\documentclass[fleqn,usenatbib]{mnras}

\usepackage{newtxtext,newtxmath}
\usepackage[T1]{fontenc}
\usepackage{graphicx}
\usepackage{hyperref}
\usepackage{graphicx}
\usepackage{lineno}
\usepackage{multirow}
\usepackage[export]{adjustbox}
\usepackage{siunitx}
\usepackage{booktabs} 
\usepackage{float}  
\usepackage{subcaption}

\DeclareRobustCommand{\VAN}[3]{#2}
\let\VANthebibliography\thebibliography
\def\thebibliography{\DeclareRobustCommand{\VAN}[3]{##3}\VANthebibliography}

\usepackage{graphicx}	% Including figure files
\usepackage{amsmath}	% Advanced maths commands

\newcommand{\hrieuv}{HRI\textsubscript{EUV}~}

\title[Nanojets associated with a prominence eruption observed with \hrieuv]{Reconnection nanojets associated with a prominence eruption observed with Solar Orbiter/EUI-HRI}

\author[Wallace \& Antolin]{Tarhik Wallace$^{1}$,
Patrick Antolin$^{1}$%\orcid{0000-0003-1529-4681}
\thanks{E-mail: patrick.antolin@northumbria.ac.uk}
\\
$^{1}$School of Engineering, Physics and Mathematics, Northumbria University, Newcastle Upon Tyne, NE1 8ST, UK
}

\date{Accepted XXX. Received YYY; in original form ZZZ}

\pubyear{\the\year{}}

\begin{document}
\label{firstpage}
\pagerange{\pageref{firstpage}--\pageref{lastpage}}
\maketitle

\begin{abstract}
Magnetic reconnection is a proposed mechanism for nanojets associated with coronal heating. We investigate the characteristics of reconnection-driven nanojets just before and during a prominence eruption using the High Resolution Imager (HRI) of the Extreme Ultraviolet imager (EUI) aboard Solar Orbiter during its perihelion on September 30, 2024. Extreme UV (EUV) images at unprecedented high spatial and temporal resolution from \hrieuv were analysed. The dimensions and propagation speeds of nanojets were estimated and used to estimate the kinetic energies. Nanojet activity was compared with GOES X-ray flux to assess its relation to flare evolution. The high spatial and temporal resolution in the EUV was found to be essential to fully capture the properties and numbers of reconnection nanojets. Approximately 120 nanojets were detected during the eruption, with 40 analysed in detail.  Nanojets exhibited lengths of $200 - 5000$~km, widths of $200 - 500$~km and durations of $2-12$~s. Instant velocities ranged from 150 km~s$^{-1}- 600~$km~s$^{-1}$ with kinetic energies reaching $1.56\times10^{27}$~erg. These nanojets are faster, longer, more energetic and more numerous compared to previous studies. We also find clear signatures of acceleration and deceleration, reflecting magnetic tension release and reach of new equilibria. Reconnection events during the eruption were found to be  more frequent and energetically intense. Pre-flare nanojet clustering indicates small-scale reconnection may precede large eruptive activity. These results suggest that nanojets also occur in fully ionised coronal plasma, playing a role in both quiescent and eruptive solar activity.  
\end{abstract}

\begin{keywords}
Sun: transition region -- Sun: corona -- Sun: activity -- Sun: filaments, prominences -- Magnetohydrodynamics -- Instabilities
\end{keywords}

%%%%%%%%%%%%%%%%%%%%%%%%%%%%%%%%%%%%%%%%%%%%%%%%%%

%%%%%%%%%%%%%%%%% BODY OF PAPER %%%%%%%%%%%%%%%%%%

\section{Introduction}
%TC:ignore

The solar corona is a very dynamic region above the Sun's chromosphere, where magnetic field interactions dominate, leading to a significant increase in the temperature. Although the chromosphere has a temperature of $\approx10^4$~K, the coronal temperature exponentially rises to $>10^6$~K with the main driver for this drastic rise still unknown. The leading theories behind this heating are based on magnetic reconnection and Alfvén waves predominantly \citep{Pontin_2022LRSP...19....1P, VanDoorsselaere_2020SSRv..216..140V}. \cite{parker1988} proposed this phenomenon could be explained by the nanoflare reconnection scenario. This is where small-angle misalignments in braided field lines of coronal loops release small amounts of energy termed nanoflares ($10^{24}$~erg), which are redistributed along the field line \citep{cargill1994}. For decades, numerical models have been used to try and isolate nanoflares observationally by detecting their rapid temperature increase \citep{Kucera_2024,cargill1994}. Observations of the solar corona in X-rays and EUV have been successful at detecting small intensity brightening events on the nanoflare range, with temperature estimates of $\sim10^7$~K \citep{Upendran2022,Ishikawa_nature_2017,Testa_2013ApJ...770L...1T} but a direct link to reconnection as a cause for these intensity enhancements, as proposed by Parker, has been hard to establish \citep[although see][for an alternate way to directly detect reconnection in action]{Testa_2014Sci...346B.315T}.

\citet{antolin2021} discovered direct evidence of reconnection-based nanoflares, known as nanojets. These nanojets are reported to be short small-scale bursts in the corona of energy on the order of the nanoflare range typically. It has been confirmed by \citet{antolin2021,Sukarmadji_2024} that nanojets have the following general characteristics: they are mostly unidirectional and transverse to the guide field, with a small bidirectional population, they have lengths ranging from $1,000 - 2,000$~km, widths of approximately 500~km, durations of 15~s or less, with speeds in the range of $100 - 200$~km~s$^{-1}$, leading to energies on the nanoflare range typically. These nanojets have been interpreted to come from small-angle reconnection (so-called component reconnection). The sudden release of magnetic tension combined with the localised heating in the small current sheets and taking into account the exposure time of the instrument leads to a jet-like dynamic blurring transverse to the guide field, i.e. the nanojet. The direction of this outflow was modelled by \citet{Pagano2021,Cozzo2025} and it was found that the braiding and curvature of the loops may play an important role in the asymmetry of the nanojets, and that the inward jets are more likely to occur and have higher energy compared to the outward jets. All observations of nanojets to date suggest the presence of braiding in the loop structures, which is expected to come from the slow photospheric granulation.

Additional research by \citet{Sukarmadji2022} emphasised that dynamic instabilities such as Kelvin–Helmholtz (KHI) and Rayleigh–Taylor (RTI), can act as key drivers of nanojets. Nanojets were found to form individually or in clusters, often near the apex or edges of curved coronal loops. Observations in diverse environments, like blowout jets and coronal rain loops, suggest that they are a general product of reconnection, largely independent of the specific instability involved. Their presence supports models of nanoflare heating in the solar corona.

Observations by \citet{patel2022} with the Coronal Imager 2.1 (Hi-C 2.1) \citep{rachmeler2019}, which has a plate scale of $0\arcsec.129$~pixel$^{-1}$ and cadence of 4.4~s, have provided high spatial and temporal resolution imaging of jet-like features interpreted as nanojets. While most show also a multi-thermal nature and some do show similar speeds and morphologies as previous reports, they also found significantly lower speeds down to 35~km~s$^{-1}$, with twice as large widths and duration. These discrepancies may be explained by the fact that the on-disk background to the analysed loops is variable, so the analysed jet-like structures may not occur in the coronal loops but actually correspond to transient events in the background (e.g. the transition region), where reconnection tends to lead to field-aligned plasma motions.

Despite these advances, open questions remain regarding nanojets. The vast majority of nanojets observed to date were detected with the \textit{Interface Region Imaging Spectrograph} \citep[IRIS,][]{DePontieu_2014SoPh..289.2733D}, mainly with the Slit-Jaw Imager in the 1330~\AA\, and 1400~\AA\, passbands, which form in cool conditions typical of the low transition region. The nanojets show rapid signatures of heating to coronal temperatures, captured in the EUV passbands of the \textit{Atmospheric Imaging Assembly} \citep[AIA,][]{aia}. Hence, it is not yet clear how common these events are, primarily because of the extreme spatial and temporal resolution required for them to be detected. Moreover, the question of whether nanojets also occur in the fully ionised solar corona is also lingering, or whether the partially ionised material such as coronal rain and prominences play a special role in facilitating nanojet formation. 

Prominence eruptions are violent displays of solar activity and are important to understand as a space weather phenomenon, as these eruptions precede flares and coronal mass ejections (CMEs) \citep{devi2021,gopalswamy2003}. These eruptions originate from unstable magnetic field configurations caused by destabilised flux ropes, where highly twisted flux ropes containing cool atmospheric gas are suspended within the solar corona \citep{chitta2025,Xia_2014ApJ...792L..38X,Vial2014}. The destabilisation process can come from magnetic reconnection or magnetohydrodynamic (MHD) instabilities like kink or torus instabilities \citep{Keppens_2019RvMPP...3...14K,Aulanier_2010ApJ...708..314A}, which propel the prominence outward, rapidly reconfiguring the surrounding magnetic fields. The expansion of these flux ropes releases large amounts of magnetic energy and mass into the solar atmosphere, causing space weather events with potential impacts on the Earth's magnetosphere.

In this paper, we look at reconnection nanojets in a new light by studying their characteristics using Solar Orbiter's \hrieuv \citep{Muller2020,EUIHRI}. This offers a unique advantage as \hrieuv provides the highest spatial resolution of the EUV spectrum to date, combined with a very high cadence, allowing unprecedented detail in both the spatial and temporal evolution of nanojets. The dataset analysed corresponds to a prominence eruption captured by \hrieuv during last year's fall perihelion. \citet{chitta2025} reports an avalanche reconnection process leading to the flare that follows the prominence eruption. The reported processes are associated with hard X-ray radiation, whose location appears to include the prominence and may be related to the nanojets. \citet{Gao2025} has provided a first look at the nanojets occurring in this eruption and found, in particular, much higher speeds up to 1000~km~s$^{-1}$ or more. {\citet{Bura_2025ApJ...988L..65B} analyse the same event and compare it to a confined event characterised by a less energetic flare, also observed by \hrieuv. Besides the high speeds matching those of \citet{Gao2025} they find much smaller speeds in line with previous reports, suggesting that the nanojet energies reflect the stored free magnetic energy in the system.} We here extend the analysis of the nanojets in the large prominence eruption, and show previously undetected unique features of the nanojets. 

In Section \ref{sec.2} we describe the event and the data used. Section \ref{sec.3} describes the methods used in this paper. In Section \ref{sec.4} we look at characteristics such as lengths, widths, durations and speeds that characterise them and compare nanojets that occur before and during the eruption. We will also note any interesting features that have been made clearer compared to previous works in this field. Section \ref{sec.5} discusses and concludes what was found in this paper.

\section{Observations}\label{sec.2}
\begin{figure}
    \centering
    \includegraphics[width=0.8\linewidth]{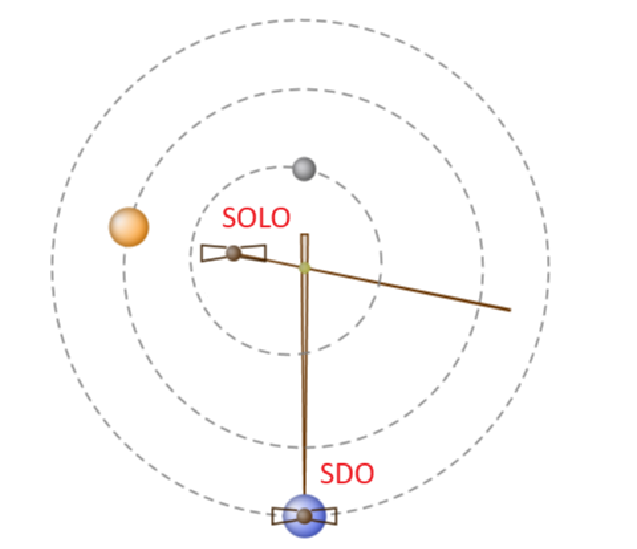}
    \caption{Location of Solar Orbiter (SOLO) and Solar Dynamic Observatory (SDO) with respect to Mercury (grey), Venus (gold) and Earth (blue) on September 30\textsuperscript{th} 2024 at 23:30:00 UT along the elliptical plane. The line-of-sight (LOS) of SOLO and SDO are indicated in the sketch.}
    \label{fig:satellite position}
\end{figure}

\begin{figure*}
    \centering
    \includegraphics[width=0.8\linewidth]{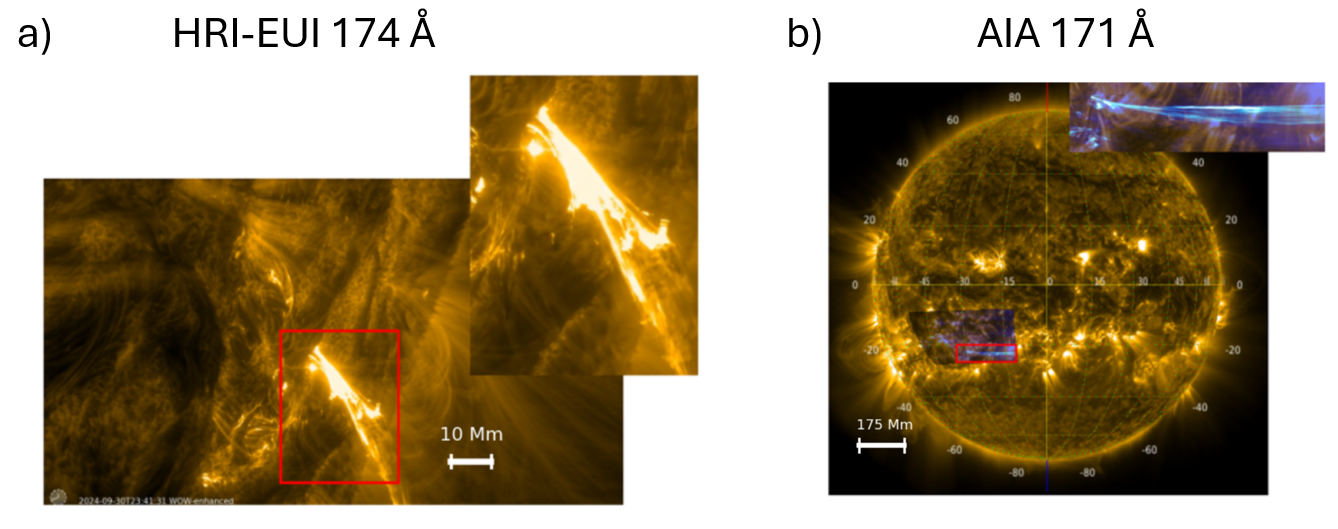}
    \caption{ (a) Coronal overview of the impulsive phase of the prominence eruption at 23:41:31 universal time (UT), observed by \hrieuv. The \hrieuv map of the eruption is displayed with the FOV of approximately 107~Mm$\times$71~Mm. The top right image is a zoom panel of the red rectangle showing the activation phase of the prominence. (b) The eruption is shown using the AIA~171~Å passband at time 23:41:00 UT with \hrieuv's FOV overlaid. The top inlet in dark blue colour shows a zoomed image of the red rectangle.}
    \label{fig:satellite image}
\end{figure*}

On September 30\textsuperscript{th} 2024, an M~7.7 class flare (SOL2024-09-30T23:47) was captured by Solar Orbiter. Because of the satellite's elliptical orbit, during its perihelion it is much closer to the Sun than any previous remote sensing satellite as seen in \citep{inbook} allowing for greater spatial resolution, meaning nanojets can be seen in greater detail. Using a propagation tool from \cite{ROUILLARD2017} the satellite can be pictured at its perihelion as shown in Figure \ref{fig:satellite position}. During this approach, \hrieuv had a spatial resolution (two pixel sampling size) of $\sim$210~km, a cadence of 2~s and an exposure time of 1.65~s. In Figure~\ref{fig:satellite image}a we show the field-of-view (FOV) of \hrieuv of a prominence eruption  (SOL2024-09-30T23:47). The event captured includes the preflare, impulsive and activation phases of the eruption. \hrieuv observes emission dominated by coronal plasma at temperatures of approximately 1~MK. Its response function is centred near 174~Å but is dominated by emissions from \ion{Fe}{IX}~171.1~\AA\, and \ion{Fe}{X}~174.5~\AA\, and 177.2~\AA\, spectral lines. Figure~\ref{fig:satellite image}b shows the view from AIA with the FOV of \hrieuv overlaid where the prominence eruption occurred. 

The prominence initially appeared as a dark, filamentary structure suspended in the corona because of EUV absorption by the neutral hydrogen and helium (and singly ionised helium) in the cooler, denser chromospheric plasma embedded within it. This filament was part of a magnetic flux rope anchored in the photosphere and surrounded by a complex system of coronal loops. Approximately 30 minutes prior to flare onset, a distinctive X-shaped magnetic configuration developed at the interface between the filament and adjacent loops, indicative of complex three-dimensional magnetic reconnection geometries \citep{chitta2025}. The filament eruption progressed into the impulsive phase marked by a steep rise in both soft and hard X-ray emissions, with reconnection appearing to occur within the erupting flux rope itself, as opposed to the classical current sheet formation beneath the rising filament. This divergence from standard flare models suggests an internal reconnection mechanism driving the flare onset \citep{chitta2025}.

\section{Methods} \label{sec.3}
Prior to analysis, the \hrieuv images were corrected for instrumental effects and jittering{, as a post process step, after} turning the data files from Level 1 to Level 2 with the euiprep routines. The regions of interest were manually selected based on observed activity and morphological consistency with previously reported nanojet structures as seen in \citet{antolin2021,Sukarmadji_2024}.

To study the evolution of nanojets during the flare event, cumulative summation of the subplots was done to see the fully developed nanojet, { as seen in Figure \ref{fig:cumulative summation}}. This was done because of the very high resolution and cadence of \hrieuv, making the nanojets appear partially developed. This differs from previous studies like \cite{antolin2021,Sukarmadji2022} where the nanojets are mostly captured during a single snapshot because of the long exposure time in those observations. The time-distance intensity maps were achieved by applying a method called bilinear interpolation to the pixel data, selecting a defined interpolation line aligned with the projected trajectory in the plane-of-the-sky of each nanojet and iterating through time. For any time-distance maps that have high intensity levels that make the nanojet hard to discern, a running difference was used by taking the current frame and subtracting it from the previous frame. This method allowed for a clearer evolution of the nanojet’s motion through the coronal environment. From this, the speed of the strand separation (bulk velocity, $v_{\mathrm{bulk}}$) was calculated by measuring the gradient of the feature in the time–distance diagram. The associated error was derived from the uncertainties in both the feature’s length and duration. To estimate the lengths of the nanojets, repeated measurements were taken of the structure, allowing for the determination of an average value along with its corresponding error. The time-distance map was used by taking the edge of the nanojet as it first appears in time and the end point of the feature as it finishes developing. The instant velocity (a better representative of the nanojet velocity) was calculated by taking the maximum extension of the nanojet in one snapshot and dividing it by the exposure time, expressed in Equation \ref{eq.1}: 
\begin{equation}
    v_{\mathrm{inst}}=\frac{x}{\Delta t},
    \label{eq.1}
\end{equation}
where the instant velocity is $v_\mathrm{{inst}}$, the exposure time is $\Delta t= 1.65$~s, and $x$ is the maximum extension of the nanojet {calculated from the time-distance diagram by selecting the snapshot of maximum extension of the nanojet. We then take a vertical slit and measure the length}. It is important to note that $v_{\mathrm{inst}}>v_{\mathrm{bulk}}$ {as $v_{\mathrm{inst}}$  represents the fastest instantaneous speed of a given nanojet}.

The widths of the nanojets were measured by taking a slit perpendicular to the nanojet and analysing the intensity profile. The width was quantified using the Full Width at Half Maximum (FWHM) of a Gaussian fit, calculated as:
\begin{equation}
    \text{FWHM} = 2 \sigma\sqrt{2 \ln(2)},
    \label{eq.2}
\end{equation}
where $\sigma$ is the standard deviation of the Gaussian {with the error being calculated using $2\sigma_{\text{err}}\sqrt{2\ln(2)}$ where $\sigma_{\text{err}}$ is the error in the standard deviation}. This measure provides a reliable estimate of the spatial confinement of the plasma outflow. Assuming cylindrical geometry, the volume of the nanojet is simply {$V=\pi r^{2}l$, where $r$ is half of the FWHM} width and $l$ is the length. The kinetic energy ($E_{k}$) can be calculated as:
\begin{equation}
    E_{\mathrm{k}}=\frac{1}{2}\rho v^{2}_{\mathrm{inst}}V
    \label{kinetic energy}
\end{equation}
where $\rho$ is the mass density. Here, we take a range of {$10^{-10}-10^{-9}~\mathrm{kg~m^{-3}}$}. This range is of commonly found values of plasma densities of prominences \citep{Vial2014}. 

In Section \ref{sec.4}, one of the points that is discussed is the the evolution of nanojets over the course of the flare using X-ray data from the GOES-16 sensor. When calculating the event timestamps, we correct the GOES data timestamps by subtracting the light travel time from the Solar Orbiter's radial distance to Earth. This ensures that all timestamps used in the paper are in Solar Orbiter’s frame of observation.

\begin{figure*}
    \centering
    \includegraphics[width=0.8\linewidth]{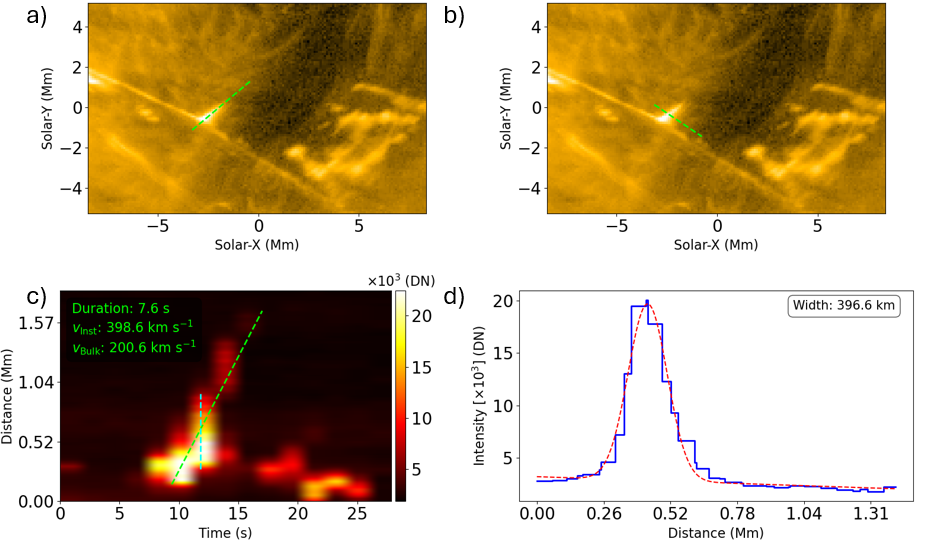}
    \caption{(a) A typical reconnection nanojet observed at 23:47:37 UT post activation of the prominence with a slit showing the direction of travel. (b) the perpendicular slit across the nanojet for the width to be measured. (c) Time-distance diagram from 23:47:26 UT to 23:47:53 UT illustrating the nanojet's evolution. The slope of the feature gives the bulk velocity. The cyan vertical dashed line shows the length used to calculate the instant velocity of the nanojet. (d) Intensity profile and the corresponding Gaussian fit along the slit shown in (b) for the calculation of the width. }
    \label{fig:typical nanojet}
\end{figure*}

\section{Results and Analysis}\label{sec.4}

The presence of nanojets is evident throughout the prominence eruption, with a clear example shown in Figure~\ref{fig:typical nanojet} where the jet-like structure is very prominent and has an instant velocity of $398.6\pm54.3~\mathrm{km~s^{-1}}$ and a bulk velocity of $200.6\pm14.3$~km~s$^{-1}$. In the time-distance diagram of Figure~\ref{fig:typical nanojet}c it is seen that the intensity brightening along the jet, following the magnetic reconnection event, fades into the surrounding environment. The width of this nanojet is found to be {$369.6\pm3.5$}~km with a length of $1524.4\pm73.1$~km. The kinetic energy is $1.5\times[10^{25}-10^{26}]\pm2.0\times[10^{24}-10^{25}]$~erg, taking the range of plasma densities considered. 

\begin{figure*}
    \centering
    \includegraphics[width=0.8\linewidth]{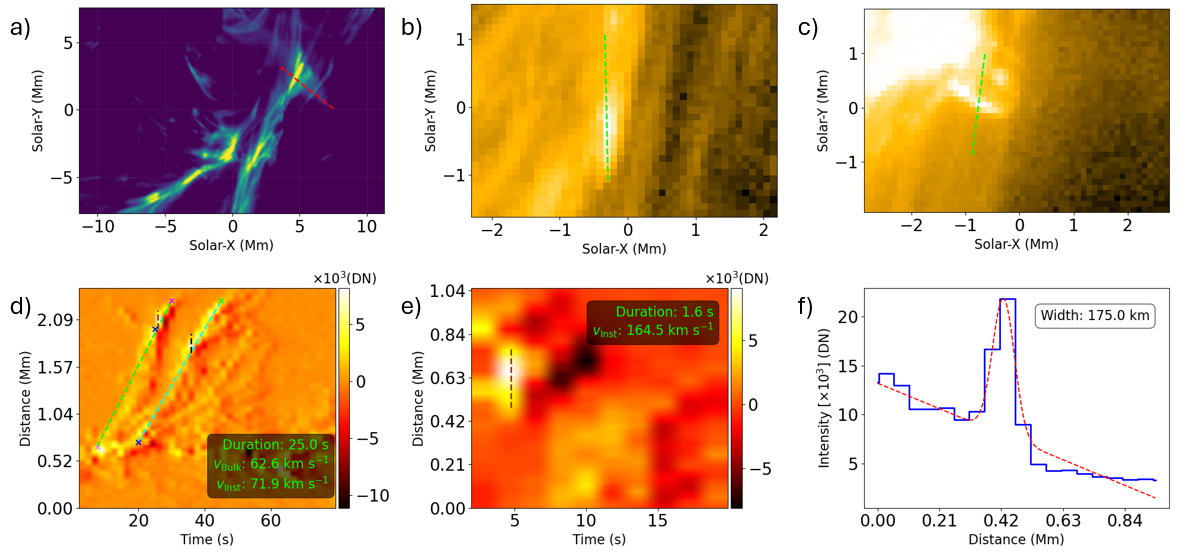}    
    \caption{(a) The longest duration nanojet captured during the activation  happening before eruption at 23:27:56 UT. (b) Nanojet with the shortest duration, observed at 23:29:35 UT.  (c) Nanojet with the thinnest width, recorded at 23:25:11 UT. (d) Time–distance diagram corresponding to (a), with speed and duration indicated in the {bottom right} corner. (e) Time–distance diagram corresponding to (b). (f) Gaussian fit to find the shortest width corresponding with (c). }
\label{fig:extremes1}
\end{figure*}

\begin{figure*}
    \centering
    \includegraphics[width=0.8\linewidth]{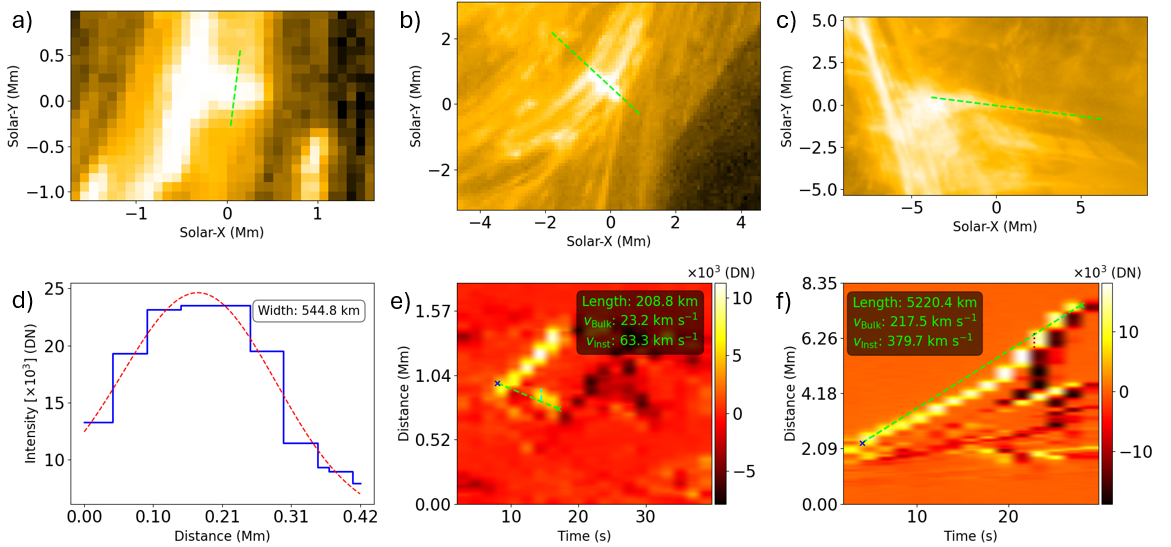}  
    \caption{
(a) Cutout showing the nanojet with the thickest width at 23:26:14 UT. (b) bidirectional nanojet with the smallest length observed at {23}:26:12 UT. (c) Longest nanojet, observed at 23:40:32 UT. (d) Intensity profile along the green dashed line in (a), with a Gaussian fit and FWHM displaying the width of the nanojet. Intensity map of (c) with characteristics shown in top left. (e) Shortest length calculated from intensity profile along with its corresponding bulk and instant velocity. (f) Data used to calculate the longest length corresponding to (e), displayed in the top left corner.}
\label{fig:extremes2}
\end{figure*}

We estimate { by visual inspection that there are at least 120 nanojets that} have been observed during the eruption by \hrieuv, representing the largest number of nanojets reported to date within a single event. However, it is likely that additional nanojets go undetected, possibly due to their small size, tight clustering and low intensities. It is also possible that some are obscured by EUV absorption, especially in the filament itself, before its activation as most nanojets are accompanied by strand separation. From the nanojets detected, 40 of them were analysed based on their clarity, spatial and temporal occurrence covering different locations and instances throughout the prominence eruption. Among the 40 nanojets analysed, there was a large variability in their characteristics. Figure~\ref{fig:extremes1} shows the extreme cases found. The longest duration nanojet (Figure~\ref{fig:extremes1}a) lasting for $25.0\pm0.4~\mathrm{s}$, the shortest nanojet (Figure~\ref{fig:extremes1}b), with a duration roughly equal to the exposure time, $1.65~\mathrm{s}$  the thinnest nanojet (Figure~\ref{fig:extremes1}c) with a width of $175.0 \pm${$3.5~\mathrm{km}$}.The thickest nanojet (Figure \ref{fig:extremes2}a) measuring $544.8\pm17.3~\mathrm{km}$, the smallest nanojet (Figure \ref{fig:extremes2}b), which was a bi-directional nanojet, with a length of $208.8\pm${$14.6$}~km. the longest nanojet (Figure \ref{fig:extremes2}c) reached a length of $5220.4\pm229.7$~km. The maximum kinetic energy found for a single nanojet has a value of $1.6\times[10^{26}-10^{27}]\pm4.1\times[10^{25}-10^{26}]$~erg. {It should be noted that in Figures \ref{fig:extremes1}a and \ref{fig:extremes2}b, overlapping strands are present. To better determine their characteristics, subtracting the intensity of the previous frame from the current frame and plotting the time-distance diagram highlights the path of the nanojet more clearly.} Most of these extremes are reported here for the first time, and highlight the highly dynamic and diverse nature of reconnection nanojets observed during the analysed eruptive event. We also note clear acceleration and deceleration patterns, best visible for the long (and longer-lived) nanojets, such as that of Figure~\ref{fig:extremes1}a. The acceleration and deceleration are seen, respectively, at the beginning and at the end of the nanojet, reflecting the work of the magnetic tension force that move the newly reconnected field lines into a new equilibrium position.

\begin{figure}
    \centering
    \includegraphics[width=\linewidth]{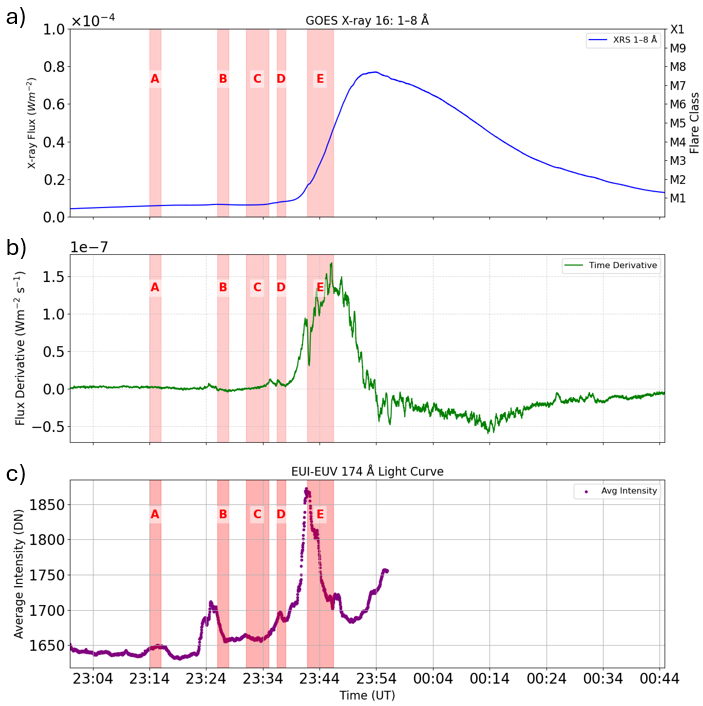}  
    \caption{(a) GOES-16 X-ray Sensor (XRS) 1–8 Å flux between 23:00 UTC on 2024 September 30 and 01:00 UTC on 2024 October 1. The curve covers the time of the prominence eruption. Times have been adjusted to match the \hrieuv time frame. The red highlighted regions labelled A–E indicate the occurrence of reconnection nanojet clusters observed by \hrieuv. The secondary $y-$axis on the right shows the standard GOES flare classification thresholds indicating that this was an M-class flare.  (b) Derivative of the GOES time-series light curve, {with highlighted nanojet clusters occurring at intervals}: \textbf{A} (23:08–23:10 UT), \textbf{B} (23:20–23:22 UT), \textbf{C} (23:25–23:29 UT), \textbf{D} (23:31–23:32 UT), and \textbf{E} (23:37–23:41 UT). (c) Temporal evolution of the average intensity over the entire FOV of \hrieuv during the prominence eruption. The shaded red regions labelled A–E indicate intervals corresponding to major clusters of nanojets.}
    \label{fig:intensities}
\end{figure}

Not only do nanojets appear as singular jet-like structures, but they also appear in clusters in small localised scales, where 2 or more nanojets can appear at the same time. This occurrence of clustering on a small scale has been reported before \citep[e.g.][]{antolin2021}, and suggests that reconnection events may occur simultaneously or sequentially (very rapidly) within regions of magnetic strands that possess similar magnetic tensions. Figure \ref{fig:intensities}~a, shows GOES-16 X-ray sensor (XRS) measurements of the 1–8~Å channel that show the evolution of the prominence eruption through the observed X-ray flux between 22:59 UTC on 2024 September 30 and 01:00 UTC on 2024 October 1.  Although many clusters of nanojets can be observed, we identify five main nanojet clusters along the observational sequence, that concentrate a large number of the analysed nanojets in this study. The red highlighted areas mark the locations of the main nanojet clusters, all of which occur prior to the onset of the impulsive phase of the flare. The first cluster appears approximately 45~min before the peak flux, with subsequent clusters occurring progressively closer in time. This temporal ordering suggests a build-up of small-scale reconnection activity ahead of the main energy release, very much in line to the avalanche effect proposed by \citet{chitta2025}. Figure \ref{fig:intensities}~b, shows the time derivative of the GOES 16 X-rays with the large clusters of nanojets highlighted over the top of the graph. It is important to note that the times of Figure \ref{fig:intensities}~a and Figure \ref{fig:intensities}~b were adjusted to match with the \hrieuv frame of reference. Figure \ref{fig:intensities}~c presents the light curve of the average intensity of the region captured by \hrieuv and shows distinct periods of increased nanojet activity (shaded intervals A--E). We notice that the EUV light curve is closely correlated with the GOES X-ray derivative, as reported already by \citet{chitta2025}. In addition we note that the occurrence times of the nanojet clusters are also correlated to some extent with the peaks of these light curves. These clusters indicate more complex magnetic topologies, where multiple small-angle reconnection events are likely triggered by the presence of complex, closely aligned magnetic field lines.

\begin{figure}
    \centering
    \includegraphics[width=\linewidth]{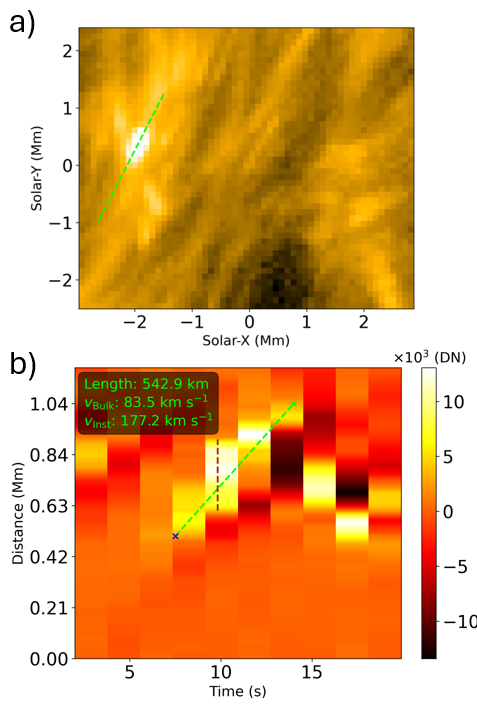}
    \caption{(a) EUV image showing the location of an oblique nanojet (highlighted by the green dashed line showing the direction of travel) within the solar corona at time 23:35:59~UT. (b) Corresponding time-distance diagram constructed along the dashed path shown in (a), illustrating the propagation of the nanojet. The green dashed line in panel (b) gives the bulk velocity and the duration. The  instant velocity is derived from the vertical brown dashed line.}
    \label{fig:oblique nanojet}
\end{figure}

It should also be noted that there were some nanojets that had an oblique line of travel, as shown in Figure \ref{fig:oblique nanojet}, which lasted for $\sim6.5\pm0.4$~s, had a length of $542.9\pm52.1$ km and a width of $301.6\pm21.8$~km, and showed an instant speed of $177.2 \pm 20.6~\mathrm{km~s^{-1}}$ with a bulk velocity of $83.5 \pm 7.3~\mathrm{km~s^{-1}}$.  \cite{Sukarmadji2022} claimed this could be a projection effect, where the nanojet is in reality closely perpendicular to the local magnetic field.

\begin{figure*}
    \centering
    \includegraphics[width=0.8\linewidth]{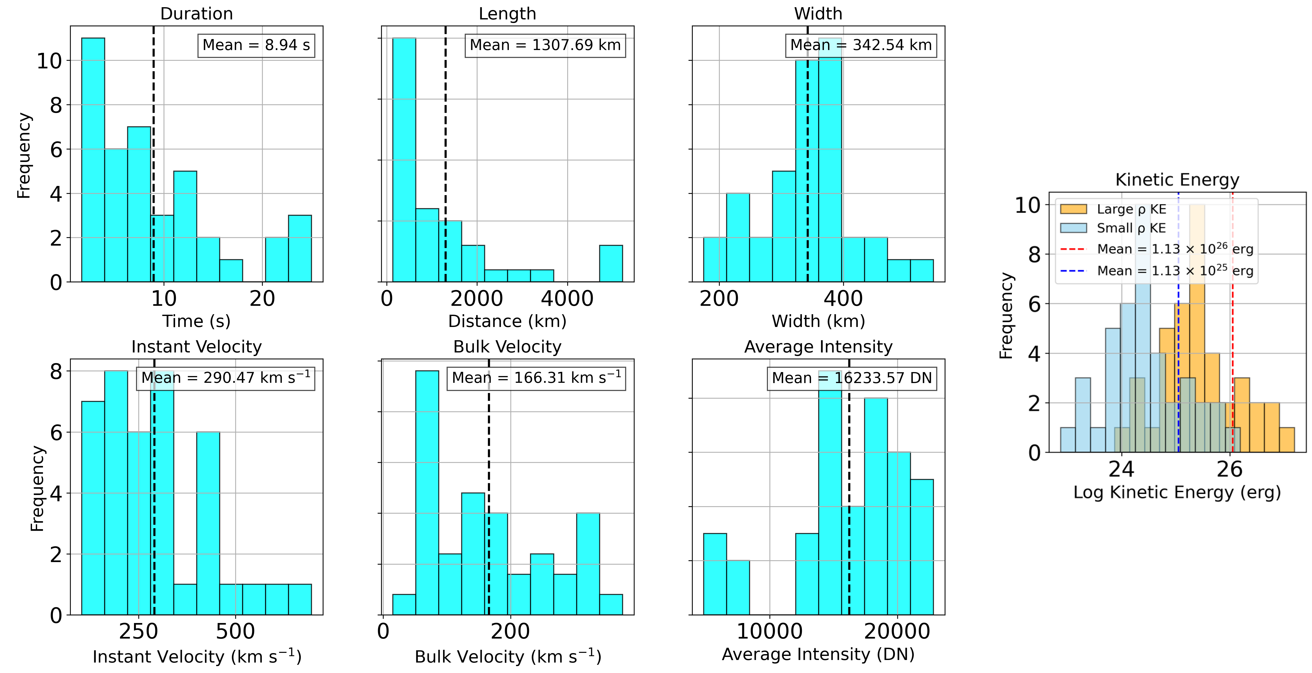}
    \caption{Histograms of the durations, lengths, widths, {instant} velocities, bulk velocities, average intensities and kinetic energies of the nanojets from all observations. With dashed lines representing the means which are displayed in the top right and top left for kinetic energy. For kinetic energy histogram, the orange bins represent a high plasma density kinetic energy and blue bins represent low plasma density kinetic energy. The red vertical line represents the high plasma density kinetic energy mean and the dark blue vertical line represents the low plasma density kinetic energy mean. }
    \label{fig:total histogram nanojets}
\end{figure*}

\begin{figure}
    \centering
    \includegraphics[width=\linewidth]{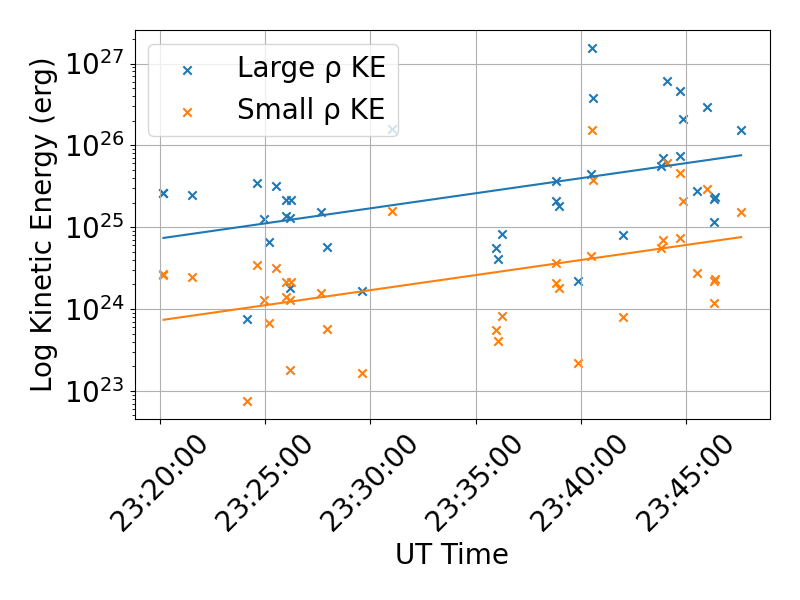}
    \caption{Kinetic energy values (in logarithmic scale) of all the nanojets over time with a line of best fit for the cases of low (in orange) and high (in blue) plasma density values. }
    \label{fig:energy time diagram}
\end{figure}

In Figure \ref{fig:total histogram nanojets} we show a histogram of all the measured parameters for all the nanojets analysed in this work. We find a mean duration of the observed nanojets of \(8.9\pm1.1~\mathrm{s}\), ranging between $\approx1.65$~s and \(12~\mathrm{s}\). The lengths averaged around \(1307.7\pm212.5~\mathrm{km}\), with most lengths being less than 1000~km with a few exceeding. These lengths are on average 30\% smaller than the typical lengths of nanojets reported with IRIS \citep{Sukarmadji2022}. We note that without summing the images over which the nanojets are observed, the lengths would result in at least a factor of 2 smaller. This reflects the effect of the fast cadence of \hrieuv, which is able to capture the evolution of the nanojets in greater detail compared to previous observations. The remaining 30\% discrepancy could be due to very rapid heating to several MK temperatures, which would limit the visibility in \hrieuv. The width measurements show an average of \(343.54\pm163.45~\mathrm{km}\) and have a range of \(300 - 400~\mathrm{km}\). Instant velocities have an average of \(290.5\pm23.0~\mathrm{km\,s^{-1}}\), with most events \(\leq 500~\mathrm{km\,s^{-1}}\). The bulk velocities, reflecting typical magnetic strand propagation speeds, exhibit an average value of \(166.3\pm15.3~\mathrm{km\,s^{-1}}\), with maximum values that are on the order of average of the instant velocities. The average intensities show values typically ranging between $15000-20000$~DN, which translates into an average of $\approx1,270~$photons~s$^{-1}$ that makes these nanojets very bright in general. 
%DN2phot = 1./7.76
The distribution of kinetic energies shows a range of \(7.4 \times 10^{22}-1.6 \times 10^{26}~\mathrm{erg}\), for a small plasma density of $10^{-10}$~kg~m$^{-3}$ and \(7.4 \times 10^{23}-1.6 \times 10^{27}~\mathrm{erg}\) for a high plasma density of $10^{-9}$~kg~m$^{-3}$. These large values reflect the result of the increased magnetic tension release from the eruption. In Figure \ref{fig:energy time diagram} we show the temporal evolution of the kinetic energies of the nanojets. We note a clear upward trend that coincides with the progression of the eruption. {However there are limitations to this graph as only 40 nanojets were analysed. A larger statistical pool is needed in order to confirm this result.}

\begin{figure}
    \centering
    \includegraphics[width=\linewidth]{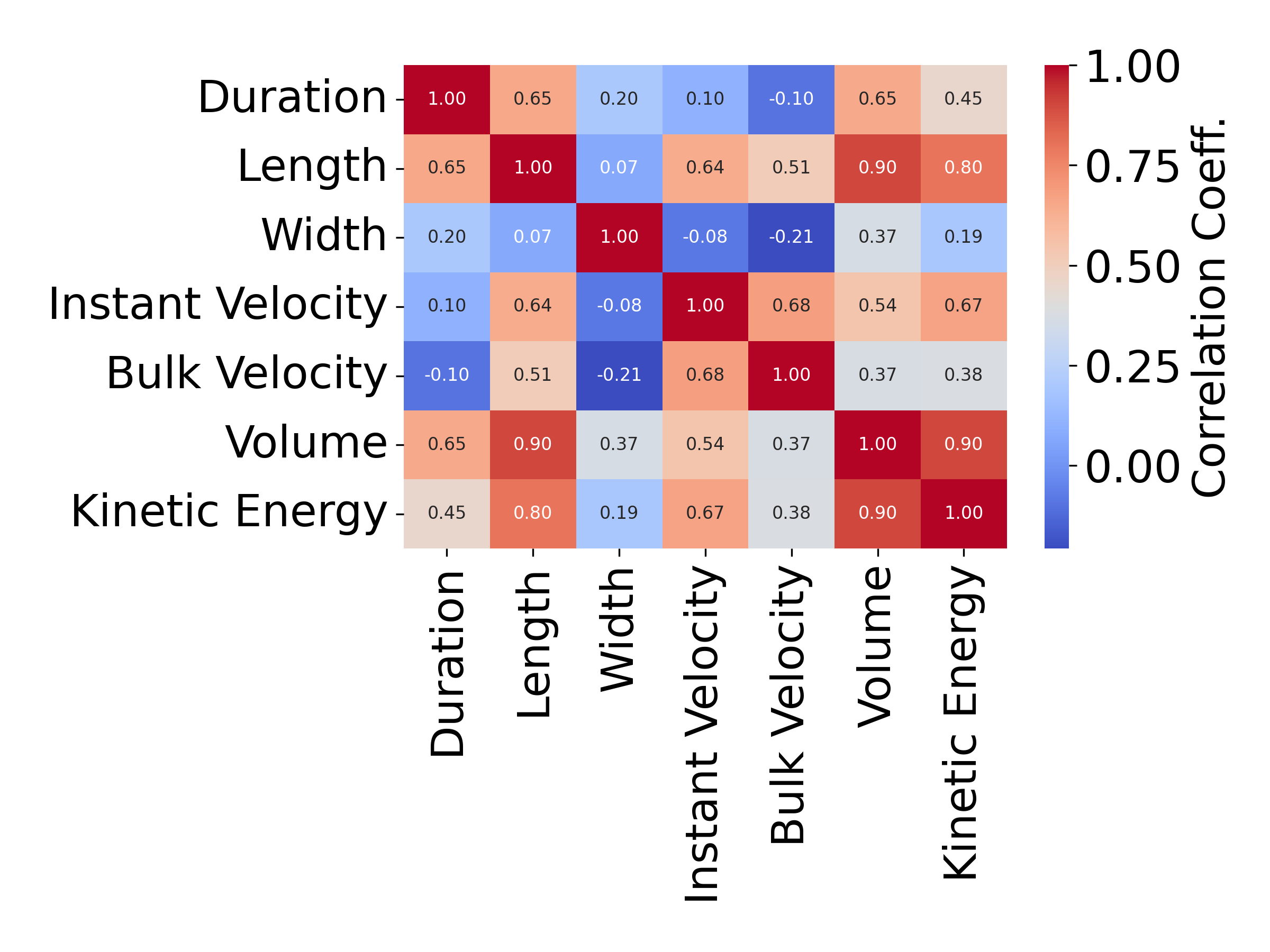}
    \caption{Heat map of all 40 nanojets showing the Pearson correlation coefficients between the duration, length, width, instant and bulk velocities, volume and kinetic energy.}
    \label{fig:correlation_heatmap}
\end{figure}

To examine the relationships between different nanojet parameters, we constructed heat maps of the Pearson correlation matrix that we show in Figure \ref{fig:correlation_heatmap}. A positive correlation is observed between the length and duration of the nanojet $r = 0.65$, which is linked to the high cadence of \hrieuv allowing to capture the nanojets being developed, and also to the way the length was calculated. We also note that the instant velocity also had a positive correlation with length with $r=0.64$. Furthermore, the instant velocity shows a strong correlation with bulk velocity $r=0.68$. Both of these are expected, since the instant velocity is strongly related to the amount of energy from the reconnection, which in turn will be reflected in the velocity of strand separation, and therefore also with the length of the nanojet. The bulk velocity shows weaker correlations with most other parameters with the only other correlation with length ($r = 0.51$). This is expected from the previously discussed correlations that imply that faster bulk motions should be associated with longer nanojets. Duration exhibits a correlation with volume with $r=0.65$. This is due to the correlation between duration and length only, because we do not find any significant correlation between duration and width. 

\begin{figure*}
    \centering
    \includegraphics[width=\linewidth]{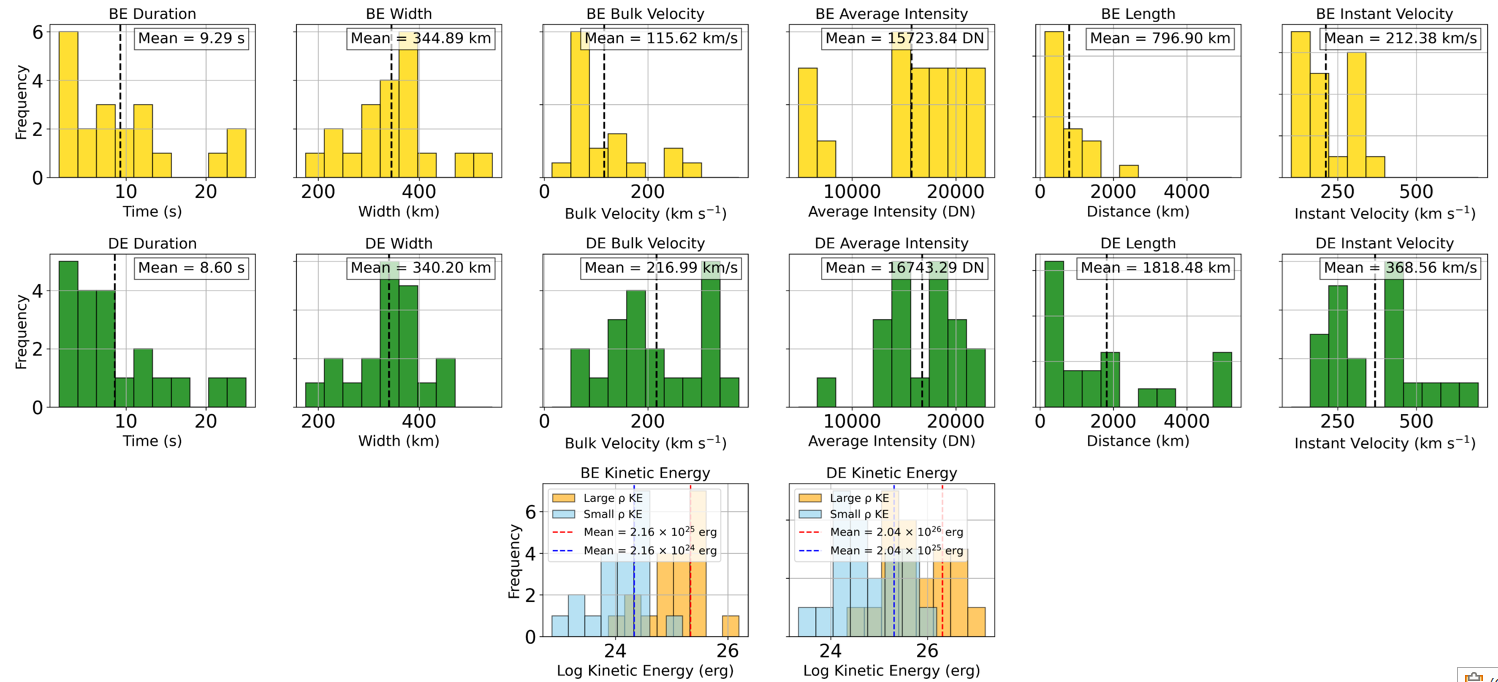}
    \caption{Histograms of before eruption (BE) and during eruption (DE) nanojets displaying durations, lengths, widths, instant velocities, bulk velocities, kinetic energies and average intensities.}
    \label{fig:nanojet histograms}
\end{figure*}

The 40 nanojets were then classified into two categories: those that occurred before the eruption (BE) and those that occurred during the eruption (DE). The motivation behind this classification was to investigate whether the onset of the eruption had any influence on the characteristics of the nanojets. Figure \ref{fig:nanojet histograms} presents a statistical comparison of nanojet properties observed before the eruption and during the eruption. Notable differences are observed between the two categories.

The mean duration of BE nanojets is $9.3 \pm 1.6$~s, whereas DE nanojets exhibit a shorter mean duration of $8.6 \pm 1.6$~s, indicating that reconnection events occur more rapidly during the eruptive phase {or that plasma is heated to higher temperatures, causing features to be visible for shorter durations in the EUI passband}. It can also be seen that DE nanojets have mean lengths and widths of $1818.5\pm375.6$~km and $340.3\pm167.3$~km, respectively, compared to $796.9\pm129.8$~km and $344.9\pm147.4$~km for BE nanojets. DE and BE widths both lie between $300-400$~km with some outliers in both categories.

The average instant velocity also increases from $212.4\pm19.1$~km~s$^{-1}$ in the BE group to $368.6\pm33.6$~km\,s$^{-1}$ in the DE group. In the BE case, the most frequent events occur between $100-370$~km\,s$^{-1}$ and for DE most events happen between $400-500$~km\,s$^{-1}$. Bulk velocities show a similar trend, with the mean rising from $115.6\pm17.2$~km~s$^{-1}$ (BE) to $217.0\pm20.0$~km\,s$^{-1}$ (DE). The highest occurrence frequency in the BE bulk velocity distribution occurs in the $25-100$~km~s$^{-1}$ range, while DE nanojets peak between $300-350$~km~s$^{-1}$, suggesting faster strand acceleration during the eruption.

The most notable difference is observed in the kinetic energy distributions. DE nanojets have a substantially higher mean kinetic energy of $2.0\times[10^{25} - 10^{26}]\pm8.1\times[10^{24}-10^{25}]$~erg, compared to $2.2\times[10^{24}- 10^{25}]\pm2.2\times[10^{22}-10^{23}]$~erg for BE nanojets. The DE group of nanojets exhibits a broader spread, with several events reaching $3.2 \times 10^{27}$~erg if a higher plasma density is assumed compared to the BE group where their kinetic energy values are fairly close together. 

These differences indicate that the eruption plays a significant role in influencing nanojet characteristics. The observed broadening in the energies, along with the increase in mean, indicates that reconnection events during eruptions are not only more frequent, but also more energetically intense. The average intensity for both groups were fairly consistent with each other with BE nanojets showing 15720~DN and 16740~DN for DE nanojets. {It should be noted that during the eruption, the filament undergoes an unwinding motion. Depending on the LOS, this may cause strong variations of the plane-of-the-sky projection of the nanojets, more in the eruptive phase than beforehand. Consequently, the observed distinction between the two phases may not be entirely physical but could be explained by projection effects. To check this, one could compare \hrieuv and AIA. Although AIA does not have the same resolution, some of the nanojets should be long enough  to be seen in both instruments. Then the total velocity can be calculated by combining both projection velocities, knowing also the angle between SolO and AIA. If the same tendency is observed with AIA between the two phases, then one can be sure that it is not a projection effect.}

\begin{figure*}
    \centering
    \includegraphics[width=\linewidth]{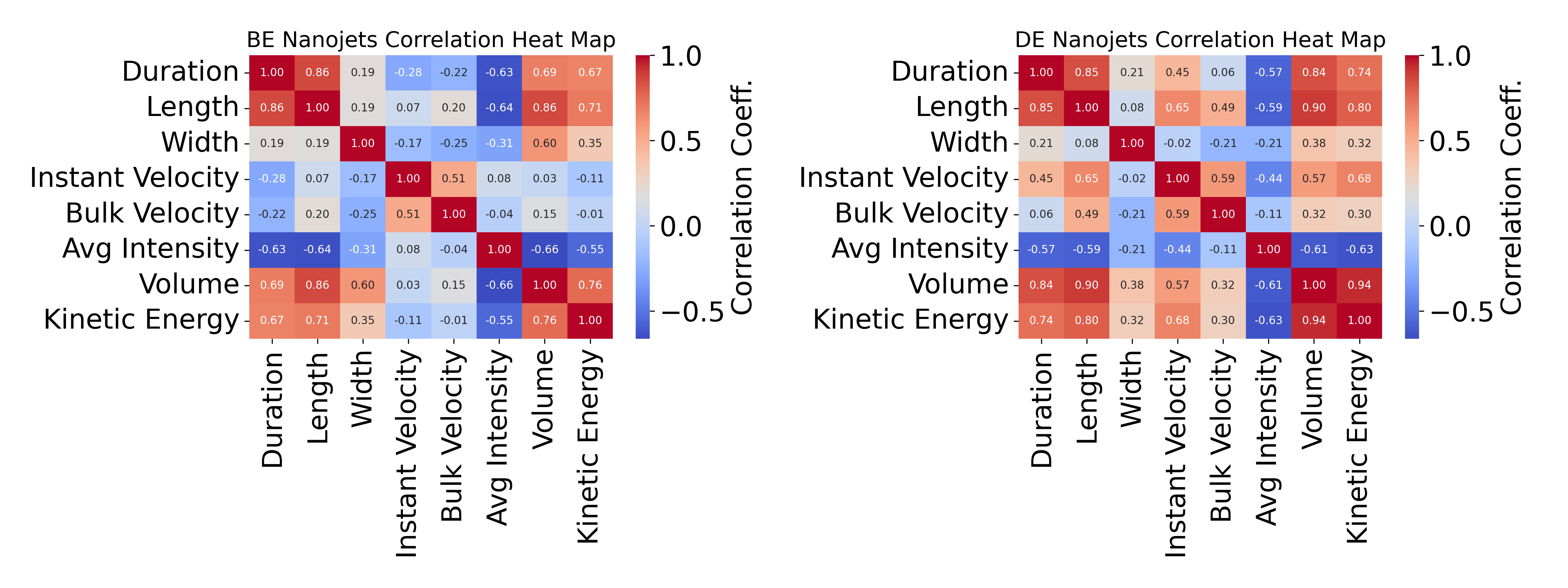}
    \caption{Pearson Correlation heat maps of BE and DE nanojets.}
    \label{fig:BE and DE heat maps}
\end{figure*}

Heat maps of nanojet parameters before and during the eruption are shown in Figure \ref{fig:BE and DE heat maps}. A clear distinction emerges between the two phases. In the BE case, correlations between most parameters are generally weak to moderate. Strong correlations are observed between duration and length ($r = 0.86$) and duration and volume ($r = 0.69$), suggesting that longer-lasting nanojets tend to extend farther and occupy more volume. Instant velocity and bulk velocity  have a moderate correlation with $r=0.5$ but for the rest of the categories they appear only weakly or negatively correlated with most other properties. 

In contrast, the DE phase shows stronger and more widespread correlations across parameters. Duration maintains strong positive correlations with length ($r = 0.85$) and volume ($r = 0.84$), while it becomes correlated to some extent with instant velocity ($r = 0.45$). Instant velocity also becomes more influential, showing increased correlation with length ($r = 0.65$).

\section{Discussion and Conclusions}\label{sec.5}

The observations presented in this study reveal several important and unprecedented findings about reconnection-driven nanojets during a prominence eruption. Previous works, including \cite{antolin2021,Sukarmadji_2024}, largely identified nanojets in cooler, partially ionised plasma such as coronal rain and chromospheric material. However, the results here using the Solar Orbiter’s \hrieuv observations during the M-class flare SOL2024-09-30T23:47, {provide evidence of strong heating to MK temperatures, which implies that the plasma becomes strongly ionised during the eruption. This raises the possibility that nanojets may also be generated within fully ionised coronal plasma under such eruptive conditions. Nevertheless, given that the filament continues to host cooler, partially ionised plasma until at least the hard X-ray peak, we cannot rule out the possibility that the observed nanojets are still rooted in cooler prominence material.}

One of the most striking findings is the identification of a large number of nanojets both before and during the prominence eruption that has not been previously recorded in this much detail with \hrieuv being able to capture the nanojet as it develops. In total, approximately $150-300$ nanojets were estimated to have occurred during the event, with 40 being analysed in detail. These nanojets show markedly different characteristics compared to those observed before the eruption. They are faster, longer and much more energetic than their pre-eruption counterparts, highlighting the influence of the evolving eruptive magnetic environment on reconnection processes. \citet{Gao2025} has provided a first analysis of the nanojets in this eruption and reported on the very fast velocities up to 800~km~$^{-1}$. {\citet{Bura_2025ApJ...988L..65B,Tan_2025arXiv250904741T} have, more recently, also investigated these jet-like features in the same event with \hrieuv, finding similarly high speeds and also favouring the nanojet interpretation.} The instant velocities from the extended number of nanojets analysed here is in line with their values. However, we note that the reported velocities by \citet{Gao2025} correspond to what we refer here as bulk velocities, which are roughly a factor of two smaller than the instant velocities. The discrepancy could be due to the relatively small pool of nanojets selected, compared to the true estimated number occurring in the same eruption. {This is supported by \citet{Bura_2025ApJ...988L..65B, Tan_2025arXiv250904741T}, who report values more in line with the bulk velocities we report.} In any case, the fact that the instant and bulk velocities can be significantly different suggests the differences between the local dynamics (the nanojet) and the global dynamics (the bulk motion of the strand separation), as well as  acceleration or deceleration of the nanojet. Indeed, in many of the time-distance diagrams we present the slope of the propagating feature is not always straight, but exhibits differences that can be associated with acceleration or deceleration (e.g. Figure~\ref{fig:extremes1}a).

Compared to previous reports we also find differences in the measured lengths of nanojets. We find that if only a single snapshot of the nanojet is measured like in \citet{antolin2021,Sukarmadji2022} (which are limited by the poor cadence) then the observed nanojet lengths are typically only a factor of two greater than their widths. This aspect ratio is notably lower than values reported in previous studies, which found ratios ranging from approximately 3 to 5. We recover these ratios when doing the cumulative summation of images. The explanation for this discrepancy is the high temporal resolution of the \hrieuv instrument aboard Solar Orbiter. 

The significantly greater number of nanojets observed in this event compared to previously studied cases suggests a possible correlation between nanojet occurrence and the total energy released during the event. Furthermore, observations show that DE nanojets have higher kinetic energies, averaging \(2.0 \times [10^{25} - 10^{26}] ~\mathrm{erg}\) compared to \(2.2 \times [10^{24} - 10^{25}]~\mathrm{erg} \) for BE nanojets (which correspond to the highest energies reported for nanojets to date), suggesting that the extreme magnetic stresses during eruptions create more powerful reconnection outflows. Accordingly, DE nanojets have a mean bulk velocity of 217.0~km~s\(^{-1}\), compared to 115.6 ~km~s\(^{-1}\) for BE nanojets, with some nanojets reaching instant velocities reaching above 600 km~s\(^{-1}\). These higher number of nanojets, more energetic in time culminating with the impulsive phase of the flare 10~min after the last main cluster, aligns with the result by \citet{chitta2025} of an MHD avalanche of reconnection events leading to major energy release.

This is further supported by the observed significant clustering of nanojets prior to the onset of the flare. This build-up of small reconnection events pinpointed by the nanojets could represent a precursor mechanism, leading to a large-scale release of energy associated with the prominence eruption. The clustering behaviour from small to large scales suggests a self-organised criticality, where multiple small-angle reconnection events may destabilise the system and trigger the larger eruption. \cite{chitta2025} found that despite the apparent random individual events, there appears to be a clear progression from initially weak to progressively more energetic reconnection episodes, as evidenced by EUV and Hard X-ray observations. This sequence is also manifested within the flux rope containing the filament, where reconnection signatures are first detected between a few threads around  23:46:00 UT and then rapidly propagate across the entire filament body in approximately two minutes. These observations, combined with the identification of nanojets, support the interpretation {of this avalanche-like self-organised criticality process where small-angle reconnection plays a central role}.

The characteristics of the nanojets analysed, show a distinctive distribution: they are generally very short-lived, with mean durations around 8.9 s, with lengths around 1307.7 km and widths around 342.54 km. These small spatial and temporal scales make them challenging to detect without the extremely high spatial resolution (\(\sim210\)~km) and cadence (2~s) capabilities of \hrieuv. The short duration and small size underline how reconnection in the solar corona can operate at extremely fine scales, contributing to energy release and plasma heating in a fragmented and intermittent manner.

Finally, the correlations found between the different nanojet properties provide further insight into the characteristics of nanojets. During the eruptive phase, strong correlations between parameters such as length, width, instant velocity, and kinetic energy suggest a more tightly coupled dynamic environment, where reconnection outflows grow in both scale and energy. Before the eruption, weaker correlations imply a more independent occurrence of reconnection events. {This picture is supported by the results of \citet{Tan_2025arXiv250904741T}, who show indications of an overlying field initially that would control and limit the reconnection. This field gets stretched throughout the eruption, offering more favourable locations for small-angle reconnection to happen.}

In conclusion, these results offer compelling new evidence that reconnection nanojets are a widespread phenomenon within both stable and eruptive coronal environments. The extreme cases observed here have not been previously reported and provide important constraints for theoretical models of coronal heating and eruption dynamics. Moreover, the large number of nanojets preceding the onset of the flare suggests that nanojet activity could serve as an important indicator of imminent large-scale solar eruptions. This study also offers tantalising first evidence that reconnection nanojets can occur within a fully ionised eruptive environment, captured during the activation of a prominence eruption. Compared to earlier studies, the nanojets observed here are more numerous, faster, more energetic, and display a broader range of sizes and durations, with a significantly higher kinetic energy than previously reported. These findings highlight the essential role that small-scale reconnection events play in the dynamics of the solar atmosphere and offer new avenues for investigating the mechanisms behind coronal heating and flare initiation.

\section*{Acknowledgements}
We would like to thank the anonymous referee for the valuable comments and suggestions, which have improved the quality of this work. Data are courtesy of Solar Orbiter and SDO. Solar Orbiter is a space mission of international collaboration between ESA and NASA, operated by ESA. The EUI instrument was built by CSL, IAS, MPS, MSSL/UCL, PMOD/WRC, ROB, LCF/IO with funding from the Belgian Federal Science Policy Office (BELPSO); Centre National d'Études Spatiales (CNES); the UK Space Agency (UKSA); the Deutsche Zentrum f\"ur Luft- und Raumfahrt e.V. (DLR); and  the Swiss Space Office (SSO). The building of EUI was the work of more than 150 individuals during more than 10 years. We gratefully acknowledge all the efforts that have led to a successfully operating instrument. SDO is a mission for NASA's Living With a Star (LWS) program.  All images in this manuscript have been made with Matplotlib on Python \citep{Hunter_2007}. This research used version 7.0 of the SunPy open source software package \citep{sunpy_community2020}.

%%%%%%%%%%%%%%%%%%%%%%%%%%%%%%%%%%%%%%%%%%%%%%%%%%
\section*{Data Availability}
The Solar Orbiter \hrieuv data used in this articles is part of Data Release 6.0 (\url{https://doi.org/10.24414/z818-4163}), which is publicly available. 
 


\bibliographystyle{mnras}
\bibliography{main.bbl}

\appendix
\section{Cumulative Summation Plot}
\begin{figure}
    \centering
    \includegraphics[width=\columnwidth]{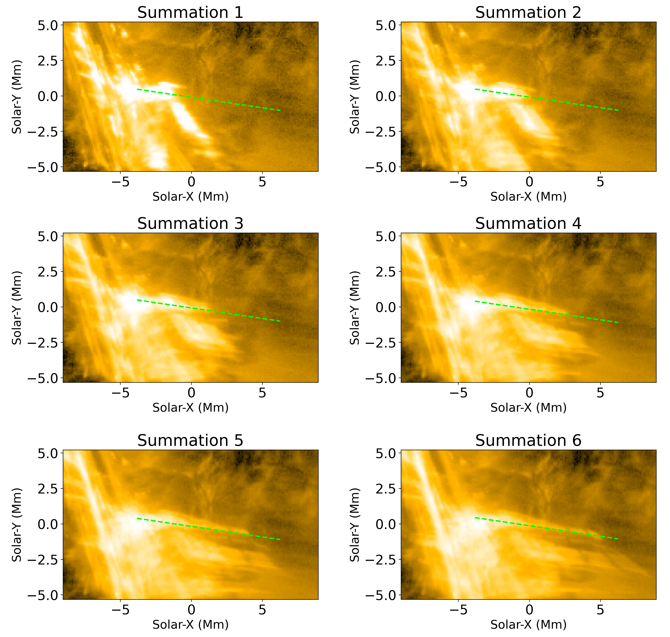}
    \caption{An example of the cumulative summation method used in identifying the full development of nanojets.}
    \label{fig:cumulative summation}
\end{figure}

To trace the temporal and spatial evolution of nanojets, a cumulative summation method was applied to the \hrieuv images. In this approach, successive frames were summed to see the nanojet fully developed as seen in Figure \ref{fig:cumulative summation}.
%TC:endignore

%\begin{thebibliography}{99}
%\end{thebibliography}

%%%%%%%%%%%%%%%%%%%%%%%%%%%%%%%%%%%%%%%%%%%%%%%%%%

%%%%%%%%%%%%%%%%% APPENDICES %%%%%%%%%%%%%%%%%%%%%

%\appendix


%%%%%%%%%%%%%%%%%%%%%%%%%%%%%%%%%%%%%%%%%%%%%%%%%%

% Don't change these lines
\bsp	% typesetting comment
\label{lastpage}
\end{document}